# Reduction in the complexity of 1D $^{1}$H-NMR spectra by the use of Frequency to Information Transformation

H. Valafar, Valafar F.

## 1. Abstract

Computer aided analysis of 1D $^{1}$H-NMR spectra is often hindered by the complexity of these spectra due to their rich information contents. Large variations that occur during the collection of these spectra further complicate the task of analyses. Large solvent and reference peaks, base line drift and negative peaks (due to improper phasing) are among some of these variations. Furthermore, some instrument dependent alterations, such as incorrect shimming, are also embedded in the recorded spectrum. The unpredictable nature of these alterations combined with the complexity of these signals have rendered the automated and instrument independent computer analysis of these spectra unreliable.

In this paper, a novel method of extracting the information content of a signal (in this paper, frequency domain 1D $^{1}$H-NMR spectrum), called the frequency-information transformation (FIT), is presented and compared to a previously used method (SPUTNIK).

FIT can successfully extract the relevant information to a pattern matching task present in a signal, while discarding the remainder of a signal by transforming a frequency domain signal into an information spectrum (IS). This technique is designed to decrease the inter-class correlation coefficients while increasing the intra-class correlation coefficients by reducing the unnecessary complexity of NMR spectra. Different spectra of the same molecule, in other words, will resemble each other more while the spectra of different molecules will look more different from each other. This feature allows easier automated identification and analysis of molecules based on their spectral signatures using computer algorithms.

## 2. Introduction

The rise of computer-assisted nuclear magnetic resonance (NMR) analysis and its applications during the past decade [1,2,3,4,5,6] has given the NMR spectroscopists new tools in the analysis of spectral signatures of large molecules. However, these tools are sensitive to small variations in the spectra and could produce false results in a noisy environment. Variation in the environmental temperature, pH level of the sample, chemical or magnetic noise, and other instrument and environment dependent fluctuations can cause alteration of peaks in NMR spectra. These alterations could effect the outcome of automated analysis of these spectra [7,8].

Several different numerical approaches have been utilized in the past to partially eliminate some of the above mentioned problems. Bayesian analysis [9,10,11,12] and polynomial fitting [13] are among some of these algorithms. The advantages and disadvantages of each of these techniques are discussed in literature review section. This section also includes a previously in-house developed technique called SPUTNIK. Section 4, discusses our newly developed FIT transformation. The data that is used in our experiments and the method of usage is discussed in Section 5, materials and methods. Section 6, discusses the results, and Section 7, is the conclusions.

## 3. Previous Methods

### *3.1. Bayesian Analysis.*

Bayesian analysis is a well studied principle and has been applied in various classification problems [14,9,10,11,12]. This analysis takes advantage of Bayes' theorem in order to calculate a-posteriori probability of an event when certain a-priori probabilities are known. One of the more famous ways of expressing Bayes theorem is shown in equation 1.

$$P(A_m|B) = \frac{P(B|A_m)P(A_m)}{\sum_i P(B|A_i)P(A_i)} \quad \text{s.t. } \exists \; i \neq j \;\; A_i \cap A_j = \varnothing \text{ and } \bigcup_{\mathrm{i}} \mathrm{A_i} = \Omega \tag{1}$$

$H_0$ and $H_1$ in the above equation can be selected to be the hypothesis that an observed intensity is noise or signal respectively [9,10]. By doing so, the points of a given spectrum can be separated into a noise or a signal class. The same analysis can be applied to baseline correction and solvent peak suppression problems by the alteration of the definitions of the hypotheses [11,12].

Bayesian classifiers, in general, tend to be very powerful and versatile. However their performance fully depends on the a-priori statistics collected about the given problem. Thus, Bayesian classifiers require either a large number of observations to draw an accurate estimation of the distribution functions or assumed knowledge about the problem. Collection of large number of data points is often impractical (especially in the case of large dimensional problems) or even not possible. That is why usually in practice, the second method of implementation of Bayesian analysis (assumptions about the distributions) is practiced. This technique, especially when combined with other techniques such as Markov processes and Monte Carlo algorithms can be very powerful [15]. Unfortunately the introduction of random mechanisms induces the disadvantages inherent to random processes. The validity of the assumptions regarding the distribution functions, convergence and the convergence time are some of these disadvantages.

### *3.2. Polynomial fitting.*

A NMR spectrum can be written as a linear combination of two components as the following:

$$\vec{S} = \vec{T} + \vec{B} \tag{2}$$

Where *S* is the acquired spectrum, *T* is the signal component of the spectrum and *B* is the drift in the baseline of the signal. This drift may occur due to several reasons such as inhomogeneity of the magnetic field or sustainability of power in the probe [16]. The *T* component of this signal can be approximated by subtracting a polynomial estimation of the baseline drift ( $\vec{B}$ ) from the acquired signal $\vec{S}$ . One disadvantage of this algorithm is that the general form of the polynomial as well as its degree is assumed ahead of time. The higher the degree of the polynomial, the more computationally intensive fitting procedure. However, higher degree polynomials could offer better fitting. Bernstein polynomials [13] are among some of the polynomials that have been used for this algorithm. These polynomials can be described as the following:

$$B(t) = \sum_{j=0}^{M} b_j t^j (1-t)^{M-j} \qquad (3)$$

The polynomial coefficients $b_j$ listed in equation 3 are calculated by minimizing the error between the speculated polynomial and the actual baseline of the spectrum. This calculation produces the second difficulty with this algorithm. Depending on the method of optimization chosen for the best fitting, different results may be obtained. One other disadvantage of this method is that it requires the baseline portion of the signal to be separated from the peaks before performing the polynomial fitting. Although this step appears to be trivial and human experts do this with ease, automated peak detection by a computer could become complicated, as most peak detection algorithms assume a reasonably well behaved baseline.

### *3.3. SPUTNIK.*

1. SPUTNIK is a NMR spectrum preprocessing algorithm (to a spectrum identification engine) developed and used at CCRC, by Dr. Jan Thomsen. This algorithm provides a signal processing tool to eliminate noise, solvent peak and baseline variations. This algorithm combines derivative and bin analyses in order to perform large peak elimination. Furthermore, this algorithm utilizes a moving window to discriminate peaks form baseline to reduce the noise level in a spectrum (for more detailed description of this algorithm please contact the authors).

Certain assumptions have been made in this algorithm that will hold under normal conditions; however they may fail under some other circumstances. Failing to meet the assumptions of the algorithm may distort a given spectrum severely. A spectrum with water suppression, for instance, would be severely distorted since the largest peak may no longer be the water peak. The second delicate assumption of this algorithm is that the largest peaks in the spectrum will also produce the largest difference points. Although in general this assumption is true, there are instances in which this assumption fails to hold.

### *3.4. Other techniques.*

Various other techniques for preprocessing of NMR signals have been introduced in the literature. However for space consideration, they can not all be reviewed in this paper. IFLAT and partial linear fit [*17,18*] are among these techniques. In general, each one of these algorithms possess some advantages and some disadvantages over alternative algorithms.

## 4. Frequency-Information Transformation

Frequency-information transformation (FIT) is a signal processing technique applicable to most signals including NMR spectra. This algorithm extracts information hidden in a noisy spectrum by comparing it to the statistics collected from an available library of previously identified spectra. By doing so, this algorithm reduces the complexity of a spectrum by extracting only the relevant information to the identification task in hand. This gives rise to the following two characteristics of FIT:

1. A larger library of spectra will ensure a more accurate extraction of relevant information.
2. Alteration of the library of spectra will alter the information extracted from the spectrum.

FIT transformation consists of two stages. All of the extreme and unnecessary peaks (such as water and standard peaks) present in a "raw" 1D $^1$H-NMR spectrum will be eliminated during the first stage. During the second stage of the transformation, all of the relevant peaks will be translated into the information space where only the "relative importance" of the peaks is calculated. It is important to note that since FIT only computes the importance of signal peaks relative to a library of spectra, it may eliminate peaks (regardless of their intensities) if they are found to be invariantly common to all spectra in the library. Similarly, peaks that exist in most spectra in the library are assigned a relatively low importance. This behavior of the FIT emerges from the underlying assumption that peaks (or features) which are present in all spectra can not be consequential in any identification or classification task; therefore they are not valuable and may be discarded. This feature allows a quick analysis of the unknown spectrum and identifies the information rich regions (hot regions) with respect to the library and identification task at hand. Thus, FIT can be used for two general purposes: elimination of irrelevant peaks and/or identification and extraction of hot regions.

The following two requirements are necessary for a successful FIT transformation.

1. A library of spectra which contains spectra of similar and dissimilar compounds (similarity/dissimilarity is determined by considering the identification task in hand).
2. Each spectrum needs to contain the same number of points spanning the same region of space (e.g. 5000 points between 1.0 and 5.5 ppm). This requirement is not as restrictive as the first one, since one can employ an interpolation algorithm to achieve equality in the size of the spectra.

The following describes the steps of the FIT algorithm. We have assumed a library of 80 1D $^1$H-NMR spectra of length 5,000 for the examples:

1. Vector normalize all spectra in the library by utilizing the definition provided in equation 4. This will ensure that the strong water peaks present in the spectra will be normalized to approximately the same magnitude. This step needs to be applied only if dominant solvent peaks are present.

$$\vec{N} = \frac{\vec{S}}{\left\|\vec{S}\right\|} \text{ where } \left\|\vec{S}\right\| = \vec{S}.\vec{S} = \sum_i s_i^2 \quad (4)$$

2. Find the minimum and maximum values for each of the 5000 channels. FIT considers each dimension of the spectra in the library to be an information channel. (i.e. we had 5000 information channels in our experiment, each containing 80 values from the 80 spectra in the library.) FIT, then, examines the 80 values in each channel, and records the highest and lowest intensities for that channel. Figure 1 illustrates the results of this step for our studies.
3. Determine the smallest value (lowest upper bound) that can be used to eliminate the water and standard peak without eliminating any of the other peaks by utilizing the graphs obtained from step 2. The value of 0.2 was selected for our experiments. Note that any smaller value may eliminate peaks that may potentially be diagnostic. This value is called maximum threshold and is used to eliminate large peaks. The intensity of all points above this level will be set to zero. Note that since only the portion of a peak that is above the threshold is eliminated, the tail ends ( portion of a peak that is below the threshold) of the eliminated peak will remain.
4. Vector normalize the resulting spectra again as defined previously in equation 4 to compensate for sample concentrations. This normalization is necessary since this step serves to equalize the signal intensities of all spectra in the library. The normalization that was performed in step 1 does not accomplish this due to the presence of very large peaks (e.g. solvent and standard). The presence of these peaks forces the normalization algorithm to mainly compensate for the variation in the intensities of these very large peaks (nearly three orders of magnitude larger than signal peaks) rather than the signal peaks.
5. Establish a histogram for each of the 5000 channels, using all 80 processed spectra, while recording the new minimum and maximum intensities for each channel. The number of bins allocated for the construction of the histograms is selected by the user according to the size of the library. 11 bins were chosen throughout our experiments. The constructed histogram can be viewed as a coarse estimation of the probability distribution function for each of the channels. Figures 2a-c illustrate some examples of these histograms.
6. The last step of the process extracts the information content from each spectrum by calculating the following probability measure for each point in a spectrum. This step produces a probabilistic measure of a particular peak uniquely belonging to a specific class of compounds. In equation 5 below, the floor operator is indicated by $\lfloor \ \rfloor$.

$$k = \left\lfloor \left(\frac{s^c - Min^c}{Max^c - Min^c}\right) \times Bins \right\rfloor \quad (5)$$

$$Inf^c = 1 - \frac{p^c(k)}{P_T^c} \quad (6)$$

$s^c$ ≡ Intensity of the c-th channel of a spectrum ($c^{th}$ data point of the spectrum).
$k$ ≡ The bin to which $s^c$ belongs.
$Max^c$ ≡ Maximum value for channel c.
$Min^c$ ≡ Minimum value for channel c.
$Bins$ ≡ Number of bins between Min & Max (11 in our experiments).
$Inf^c$ ≡ Information content of channel c of a given spectrum.
$p^c(k)$ ≡ Population of the $k^{th}$ bin of channel c.
$P_T^c$ ≡ Total population of channel c (equal to 80 in our experiments).

# 5. Materials and Methods

We have recorded and constructed a database of 80 1D $^1$H-NMR spectra of 23 complex carbohydrates (N-linked oligosaccharides) to test the performance of FIT. The frequency and frequency-information (produced by FIT) spectra of these carbohydrates are used to train and test ANNs. In this section we describe the recording conditions of the spectra, as well as the training and testing parameters selected for ANNs that were employed in all experiments reported in this manuscript.

## *5.1. Collection of NMR spectra.*

80 1D $^1$H-NMR spectra of 23 N-linked oligosaccharides constituted the library of spectra used throughout the experiments presented in this paper. 19 of the 23 structures were represented in the library by 4 spectra (different acquisitions of the same compound with various concnetrations), while the remaining 4 were represented by only one spectrum each, giving rise to a total of 80 spectra. Table 1 lists the CarbBank[1] [19] structure identification numbers (SINs) for the compounds in the library in addition to the number of spectra of each compound present in the library.

The spectra were collected on a Bruker AMX500 (500MHz field intensity) NMR instrument at 25°C and neutral pH. Signal-to-noise ratio of each spectrum was enhanced by performing signal averaging. Different number of transients were used for signal averaging in order to improve signal-to-noise ratio to an acceptable minimum level. Each spectrum contained at least 8192 points spanning 0.0 to 6.0 ppm. These spectra were processed by an interpolation routine to standardize their lengths (number of points). After this step, each spectrum contained 5000 points spanning 1.0 to 5.5 ppm. No water suppression was performed on any of the spectra, thus the water peaks were in general at least 3 orders of magnitude larger than a typical carbohydrate signal.

[1] Due to the complexity of N-linked oligosaccharide structures the IUPAC names of these compounds are fairly large. Therefore, the CarbBank structure identification numbers (SINs) are used as the means of referring to them without the use of their IUPAC name.

**Table 1. Table of N-linked oligosaccharides included in our experiments, and the number of spectra available for each one in our databases.**

| CarbBank SINs | Multiplicity | CarbBank SINs | Multiplicity | CarbBank SINs | Multiplicity |
|---|---|---|---|---|---|
| 17833 | 4 | 19396 | 4 | 14682 | 1 |
| 19346 | 4 | 19840 | 1 | 12752 | 1 |
| 15969 | 4 | 19793 | 4 | 17742 | 4 |
| 16574 | 4 | 16032 | 4 | 17736 | 4 |
| 18380 | 4 | 18089 | 4 | 20164 | 4 |
| 9440 | 4 | 18258 | 4 | 21344 | 4 |
| 21615 | 4 | 17321 | 4 | 19046 | 4 |
| 21068 | 1 | 16179 | 4 | | |

### *5.2. Testing procedure for FIT:*

Raw NMR data, SPUTNIK processed data and FIT processed data were used in conjunction with three different classification methodologies. Examination of the performance of each of the data sets in conjunction with a given classification method would reveal the effectiveness of each transformation in reducing the complexity of the problem. All models were developed to perform the same identification task, namely compound identification for a library of 23 compounds mentioned earlier.

The first of the three pattern recognition methodologies used, involved the use of correlation coefficient matrices as the measure of distances to perform cluster analysis. The second method applied Bayesian classification to the correlation coefficients while the third method utilized Artificial Neural Networks (ANN).

### *5.3. Artificial Neural Network (ANN) parameters:*

Several options are selected and maintained for the design and operation of all ANN, throughout all of the experiments. The most common neural network architecture, namely, two-stage, fully connected, feed-forward topology with a back-propagation learning algorithm [20] was used for the experiments reported in this manuscript. We used a step size of 0.01 with no momentum term. The topology of the network consisted of 5000 input neurons, 10 hidden neurons and 23 output neurons. Each learning session was terminated after 100,000 epochs of learning. The input to the network consisted of 5000 points in a 1D $^{1}$H-NMR spectra spanning the region 1.0-5.5 ppm. The highest output of the network determined the corresponding classification results.

The simulation software was written in-house with all standard options of neural network simulation implemented. All of the simulations were conducted on a DEC-ALPHA 2100 system with 4, 300 MHz. EV5 ALPHA processors. The simulation for each of the experiments were repeated four times and the final result consisted of the average of the results of four separate runs. The averaging attempts to reduce the effect of initial random weight distribution on the learning of ANNs.

## 6. Results and Discussion

In this section, we evaluate FIT's performance first by visual inspection, and then numerically using three different methods. Figures 3 and 5 illustrate 1D $^1$H-NMR spectra of the N-linked oligosaccharides identified by their CarbBank numbers 15969 and 19396 respectively. Figures 4 and 6 illustrate the reproducibility properties of FIT. Figure 4 displays the FIS (frequency-information spectrum) obtained from FIT transformation of two different frequency spectra of compound 15969 while Figure 6 illustrates two different FIS spectra of 19396. The two spectra plotted in each of the figures are similar enough that when superimposed, they are difficult to distinguish. This similarity indicates reproducibility of FIT results which will be further examined via numerical techniques later in this section. Figure 7 illustrates the differences of the FIS spectra of different compounds. This figure compares two FIS spectra obtained from compounds 15969 and 19396. The differences produced by the FIT transformation of different compounds contributes to the separation of classes of compounds and therefore decreases the inter-class correlation coefficients.

As it was mentioned before, FIT can be utilized for two different purposes. The first is to improve the quality of the signal by removing unnecessary information (redundant or not relevant information to the identification task) contained in a 1D $^1$H-NMR spectrum. The second is to determine the "information rich" regions of a signal relative to a given library of spectra and relative to a given identification task. This feature is more useful for the study of new and unknown signals where the information content of the signal is not yet known. We tested the second feature of FIT by examining its performance on a library of 23 well studied N-linked oligosaccharides.

It is well documented that the anomeric (5.5 to 4.5 ppm) and the hump (4.3 to 3.3 ppm) regions of 1D $^1$H-NMR spectra of polysaccharides are information rich regions of these spectra. This conclusion was confirmed by FIT as most peaks in the frequency-information domain appeared to be in these regions (see figures 4,6).

The following three sections are concerned with the numerical evaluation of FIT's performance. It is important to mention that more rigorous evaluation is in progress, and will be reported in a follow up article. The results reported here are only valid for the mentioned methods.

### *6.1. Correlation coefficient matrix as a tool to evaluate performances:*

An ideal transformation would be one that enhances the similarities of different spectra of a compound while increasing the dissimilarities between spectra of different compounds significantly. Therefore, this transformation would increase the correlation coefficients between spectra of the same compound (intra-class) and decrease the correlation coefficients between spectra of two different compounds (inter-class). Any such ideal transformation should ultimately produce a correlation coefficient of 1 for any two spectra of the same compound while producing a correlation coefficient of 0 for spectra of two different compounds. Equation 7, demonstrates an ideal correlation coefficient matrix for a library that consists of 5 spectra of two structures. The first two spectra belong to the first structure while the remaining 3 belong to the second structure. The entity $C^S$ is also introduced in equation 8 as the set of numbers that are contained within the C matrix. Different subsets of this set will be extracted and utilized for several different measurements in future sections.

$$C = \begin{array}{c} \\ S_1^1 \\ S_2^1 \\ S_1^2 \\ S_2^2 \\ S_3^2 \end{array} \begin{array}{c} \begin{array}{ccccc} S_1^1 & S_2^1 & S_1^2 & S_2^2 & S_3^2 \end{array} \\ \begin{bmatrix} 1 & 1 & 0 & 0 & 0 \\ 1 & 1 & 0 & 0 & 0 \\ 0 & 0 & 1 & 1 & 1 \\ 0 & 0 & 1 & 1 & 1 \\ 0 & 0 & 1 & 1 & 1 \end{bmatrix} \end{array} \qquad (7)$$

$$C^S = \left\{ C_{ij}^S \middle| C_{ij}^S = C_{ij} \right\} \qquad (8)$$

$S_j^i$ denotes the j-th spectrum of compound i in the database.

Although a theoretically perfect transformation would produce only 1's and 0's for the intra and inter class distances respectively, a practically prefect transformation (deviation from ideal, yet perfectly functional), on the other hand, would produce numbers close to 1 and 0 for the intra and inter class distances respectively with a distinct gap between the two. Therefore, since the deviation form a theoretically perfect transformation does not necessarily imply a non-functional transformation, it becomes necessary to develop the means for determining the efficacy of any such transformation. The following equations are used to calculate and compare the performances of different algorithms utilized in this paper. Most of these calculations attempt to find a distance between the theoretical and measured correlation coefficient matrices. Table 2 contains the verbal definition for some of the newly introduced variables.

$$n = \sum_{i=1}^{m} n_i \tag{9}$$

$$\hat{C}^S = \left\{ \hat{C}_{ij}^S \middle| \hat{C}_{ij}^S = \hat{C}_{ij} \right\} \tag{10}$$

$$Intra_m = \left\{ (i, j) \middle| \sum_{k=1}^{m-1} n_k < i, j \le \sum_{k=1}^{m} n_k \right\} \tag{11}$$

$$Inter_m = \left\{ (i, j) \middle| (1 \le i, j \le n) \wedge (i, j \notin Intra_m) \right\} \tag{12}$$

$$Intra = \bigcup_{i=1}^{m} Intra_i \tag{13}$$

$$Inter = \bigcup_{i=1}^{m} Inter_i \tag{14}$$

$$S_{Intra} \in C^S = \left\{ C_{ij}^S \middle| i, j \in Intra \right\}, \hat{S}_{Intra} \in \hat{C}^S = \left\{ \hat{C}_{ij}^S \middle| i, j \in Intra \right\} \tag{15}$$

$$S_{Inter} \in C^S = \left\{ C_{ij}^S \middle| i, j \in Inter \right\}, \hat{S}_{Inter} \in \hat{C}^S = \left\{ \hat{C}_{ij}^S \middle| i, j \in Inter \right\} \tag{16}$$

$$D_{Total} = \sum_i \sum_j (c_{ij} - \hat{c}_{ij})^2 \tag{17}$$

$$D_{Inter} = \sum_{(i,j) \in Inter} (c_{ij} - \hat{c}_{ij})^2 \tag{18}$$

$$D_{Intra} = \sum_{(i,j) \in Intra} (c_{ij} - \hat{c}_{ij})^2 \tag{19}$$

$$D_{avg} = \frac{D_{Intra}}{\|Intra\|} + \frac{D_{Inter}}{\|Inter\|} \tag{20}$$

$$\widehat{C}_{X,Y} = \frac{\mathrm{cov}(X, Y)}{\sigma_x \sigma_Y} \tag{21}$$

$$\mathrm{cov}(X, Y) = \frac{1}{n} \sum (x_i - \mu_X)(y_i - \mu_Y) \tag{22}$$

$$\sigma_X = \sqrt{\frac{n(\sum x^2) - (\sum x)^2}{n(n-1)}} \tag{23}$$

**Table 2. Definition of some used variables.**

| | |
|---|---|
| $n_m$ | Number of spectra of compound m present in the library. |
| $n$ | Total number of spectra (80 spectra). |
| $C$ | The ideal correlation coefficient matrix. |
| $\hat{C}$ | The calculated correlation coefficient matrix. |
| $c_{ij}$, $\hat{c}_{ij}$ | i,j th elements of the respective matrices. |
| $C^S$, $\hat{C}^S$ | Sets containing the elements of C or $\hat{C}$. |
| $Intra_m$ | Set containing the i,j coordinate of the Intra correlation coefficients of compound m. |
| $Inter_m$ | Set containing the i,j coordinate of the Inter correlation coefficients of compound m. |
| $Intra$ | Set containing the i,j coordinate of all Intra correlation coefficients. |
| $Inter$ | Set containing the i,j coordinate of all Inter correlation coefficients. |
| $D_{Total}$ | Total distance measured as defined in equation 17. |
| $D_{Intra}$ | Distance measured between the members of the elements of the Intra sets. |
| $D_{Inter}$ | Distance measured between the members of the elements of the Inter sets. |
| $D_{avg}$ | Weighted average of $D_{Intra}$ and $D_{Inter}$ distances. |
| $\|A\|$ | The number of elements in the set A. |
| $Cov(X,Y)$ | Covariance between two vectors X and Y. |

Table 3 lists the calculated deviation values of raw NMR, FIT processed and SPUTNIK processed spectra. Note that although the values $D_{Inter}$, $D_{Intra}$ and $D_{Total}$ are each listed separately in this table, it is a mistake to use any of these measurements by themselves to determine the overall performance of any of the transformations. Based on the $D_{Total}$ results listed in Table 3, the best classification results should be obtained by using the raw spectra without any processing. This conclusion is flawed and this error can be explained by decomposing $D_{Total}$ into its two different components. The best way to explain this concept is by producing a numerical example. 80 spectra representing 23 distinct N-linked polysaccharides constructed our database. 19 of these compounds were represented by 4 distinct spectra while the other 4 were represented with only one spectrum each. 80 spectra will give rise to 6400 (80x80) number of elements in the correlation coefficient matrix. Therefore, in calculating the total distance, $D_{total}$, 6400 pieces of information are used. However, the given 6400 pieces of information consist of two sets of numbers, namely the Intra and Inter class distances. The Intra class set contains 308 (19x4x4+4x1) elements while the Inter class set contains 6092 (6400 - 308) elements.

Given the above information, consider a transformation that produces random outputs. This transformation would take an NMR spectrum as its input and produce a random set of numbers at the output as the processed spectrum. For the following transformation, the correlation coefficient matrix should theoretically contain all zeros. The difference between this correlation coefficient and the ideal correlation coefficient matrices would exist in only the Intra distances ($D_{Intra}$ are the only non-zero numbers).

For a given true random transformation the following holds :

$$\hat{c}_{ij} = 0, \forall i, j$$

$$\therefore D_{Inter} = \sum_{i=1}^{6092} (0-0)^2 = 0 \ , \ D_{Intra} = \sum_{i=1}^{308} (1-0)^2 = 308$$

$$\therefore D_{Total} = D_{Inter} + D_{Intra} = 308$$

This means that if $D_{Total}$ is used as the ultimate performance measure, the total randomization of data would produce a better performance than any other technique listed in Table 2. Therefore, $D_{Total}$ should not be used as a measure of performance by itself. Furthermore, it is easy to rationalize why neither the $D_{Intra}$ nor $D_{Inter}$ can be a good measure of performance by themselves. Average distance ($D_{avg}$) was introduced in order to produce one coherent measure of the performance since both $D_{Intra}$ and $D_{Inter}$ are important in determining the overall performance of any transformation. $D_{avg}$ combines these two measurements by performing a weighted average of the $D_{Intra}$ and $D_{Inter}$ measurements. The following equations were used in calculating $D_{avg}$.

$$D_{avg} = \frac{D_{Inra}}{\|Intra\|} + \frac{D_{Inter}}{\|Inter\|}; \quad \|Intra\| = 308, \ \|Inter\| = 6092$$

**Table 3. Comparison .**

| Data | $D_{Intra}$ | $D_{Inter}$ | $D_{Total}$ | $D_{avg}$ |
|---|---|---|---|---|
| Raw NMR spectra | 65 | 1092 | 1157 | 0.39 |
| FIT of processed NMR spectra | 30 | 1475 | 1505 | 0.34 |
| SPUTNIK processed NMR | 15 | 2355 | 2370 | 0.44 |

It is very important for one to keep in mind that correlation coefficient analysis is a measure of the linear behavior of data and hence the correct interpretation of the results listed in Table 3 is as following: FIT can be recognized as the transformation that produces the best linearly-separable results. Since correlation coefficient analysis examines the performances in a linear sense, therefore the effectiveness of FIT in conjunction with a nonlinear classification method still remains unanswered.

In this section we have shown that correlation coefficient analysis can utilize the FIT produced data more effectively than the SPUTNIK produced data. In a later section, we will demonstrate that a nonlinear classification technique such as ANN also benefits more from the FIT transformation rather than the SPUTNIK transformation.

### *6.2. Bayesian classifier as a measure of performance:*

Bayesian classifiers [21,22,23,24,25] are well studied tools for classification purposes. These classifiers can be used in order to measure the performance of a preprocessing algorithm for NMR spectra. Ultimately, the discovery of a transformation that can be used in combination with Bayesian classifiers (or any other classifier) to achieve a higher classification performance is desirable. Therefore, we considered the transformation that produces the more accurate classification in combination with Bayesian algorithm a more effective transformation.

Bayesian classifiers require a-priori knowledge of the probability distribution function of the phenomenon in question, in order to estimate the a-posterior probability of a different but yet related phenomenon. This algorithm is based on the Bayes theorem that can be formulated in different ways. Equations 24 and 25, describe the basic decision making policy. $P(H_i|\vec{x})$ is the probability that hypothesis $H_i$ is true given that $\vec{x}$ is observed. Similarly $\hat{f}_{\vec{X}}(\vec{x}|H_i)$ indicates the probability distibution function of $\vec{x}$ given $H_i$.

$$P(H_0|\vec{x}) \underset{H_1}{\overset{H_0}{\gtrless}} P(H_1|\vec{x}) \tag{24}$$

$$\hat{f}_{\vec{X}}(\vec{x}|H_0)P(H_0) \underset{H_1}{\overset{H_0}{\gtrless}} \hat{f}_{\vec{X}}(\vec{x}|H_1)P(H_1) \tag{25}$$

The probability distribution of some of the features of the NMR spectra need to be estimated since they are not know. Various methods can be used for the estimation of a probability distribution function [14]. Histogram or bin analysis [26] are two such method. Figures 8a-c illustrate the estimated distribution functions of the intra and inter class correlation coefficients for raw data, SPUTNIK, and FIT processed spectra respectively using histogram analysis. According to Bayes estimators (classifiers), the best classification can be obtained by selecting the intersection of the two distributions as the threshold for classification. After doing so, the probability of error using correlation coefficients in combination with a Bayes classifier on raw data, SPUTNIK processed, and FIT processed spectra are indicated within Figures 8a-c.

The estimated probability of error when using the raw data is 55% while the error is reduced to 43% and 33% for using SPUTNIK and FIT processed spectra respectively. Note that although each one of these transformations improve the classification performance, they do not however produce acceptable results. This is due to the fact that the classification space is non-linear and by using a linear method we can not achieve acceptable results.

### *6.3. Nonlinear separability of data in Information Space Versus Frequency Space:*

Correlation coefficient analysis, as mentioned before, is a linear measure of the separation of data. Therefore, while the achievement of desirable results by the use of correlation coefficient is useful, the failure to do so is not necessarily unproductive. It is possible for a transformation to produce highly non-linearly-separable space where a nonlinear classification method will have a perfect performance while any linear method (such as correlation coefficient) produces discouraging results. Based on this piece of information one can set forward the argument that although FIT can improve the linear separability of a problem to some extent, the need was never there since the problem was easily separable by the use of some nonlinear technique. Utilization of ANNs as a method of nonlinear classification can help to provide some answers to the above scenario. If ANNs succeed in learning to isolate the raw NMR spectra of the compounds, then there would be no need to develop any other transformations.

Artificial neural network with a topology of 5000 input, 10 hidden and 23 output neurons was selected for this portion of the experiment. The input to this network consisted of 5000 points of a spectrum while the 23 output neurons determined the identity of the compound in question. The network was trained to produce zeros in all output neurons except 1. The appropriate neuron signifying the correct output was trained to produce an output of 1 while the remaining 22 neurons produced outputs of 0. This network remained unaltered during the experimentation with each data sets.

Figure 9, contains the learning curves for the raw NMR spectra, SPUTNIK processed spectra and FIT transformed spectra. The maximum training error is displayed in this graph as a measure of ANN's learning progress. In this figure, the maximum error for the raw NMR spectra does not decline below 0.8 after 100,000 iterations indicating the inability of a 10 hidden neuron in learning the raw data. However, the SPUTNIK processed spectra, as well as the FIT transformed spectra are learned by the ANN. The difference in learning appears in the ease and speed of learning. FIT transformed spectra were learned by the same network approximately five times faster. After 100,000 epochs of training the maximum bit error for the SPUTNIK processed arrives at approximately 0.3, while the same error is obtained in less than 20,000 iterations with the use of FIT transformation. These experiments were conducted using the identical parameters and 10 hidden neurons. While ANN failed to separate compounds by the use of their SPUTNIK processed NMR spectra with using 10 hidden neurons, FIS spectra (frequency-information spectra) were easily learned by the ANN with as little as 8 hidden neurons. The speed of the decline in the error combined with the reduction in the size of the hidden layer illustrate the effectiveness of FIT in reducing the complexity of a given data set.

## 7. Conclusion

NMR spectroscopy can be a powerful tool for composition and linkage analysis of biomolecules. However, the analysis of the collected spectra requires highly trained individuals and is therefore often costly and time consuming. The analysis of these macromolecules can be aided by the utilization of computers and pattern recognition techniques in order to reduce the cost and analysis time. One such example of computer assisted identification of biologically active molecules can be found at the web page of the Complex Carbohydrate Research Center (CCRC, http://www.ccrc.uga.edu).

1D $^{1}$H-NMR spectroscopy has been successful in the analysis of small to medium size molecules. One of the reasons of the limitations of 1D spectroscopy in analysis of large molecules is due to the information-rich content of these spectra. Often the complexity of 1D spectra can be reduced by performing multidimensional spectroscopy where the information is separated into different dimensions. Thus the information present in any one or two dimensions can be considered simplified and easier to analyze. However the implementation of multidimensional spectroscopy requires a significant increase in the acquisition time as well as the expertise of the operator. Furthermore, it can be argued that not all of the information present in a 1D spectrum is required to perform any given analysis task. For example, for the monosaccharide decomposition analysis of an oligosaccharide, it is very likely that the examination of the anomeric peaks will suffice in successfully completing this task. However it is unlikely that during the task of linkage analysis these peaks serve any function. Peaks that appear in the hump region of the spectrum are more likely to contribute toward the discovery of the linkages. It is important to keep in mind that a 1D spectrum contains information regarding electronic, atomic and molecular levels of a given molecule. Therefore, these spectra can be considered extremely rich in information that may not be useful in performing any one single analysis. The presence of additional information contributes toward the complexity of these spectra and thus the difficulty in their interpretation. The presence of these extra information make the task of performing a specific analysis difficult for both humans and computers. Therefore, the computer aided analysis of these spectra could be simplified and improved by the applications of transformations such as FIT.

In This paper, we have demonstrated the ability of FIT in reducing the complexity of 1D $^{1}$H-NMR spectra in the task of compound identification. It is important to realize that FIT is in its infancy stages and we hope that by minor improvements it can be used in much more sophisticated classification tasks. We speculate that through the use of FIT, we may be able to develop identification routines which may be able to identify complex oligosaccharide structures that are present in our database, but even furthermore be able to provide monosaccharide and linkage information for an unknown compound that is not present in our database. This way, not all of the complex oligosaccharides in existence need to be present in the training database in order to have an effective search or identification algorithm.

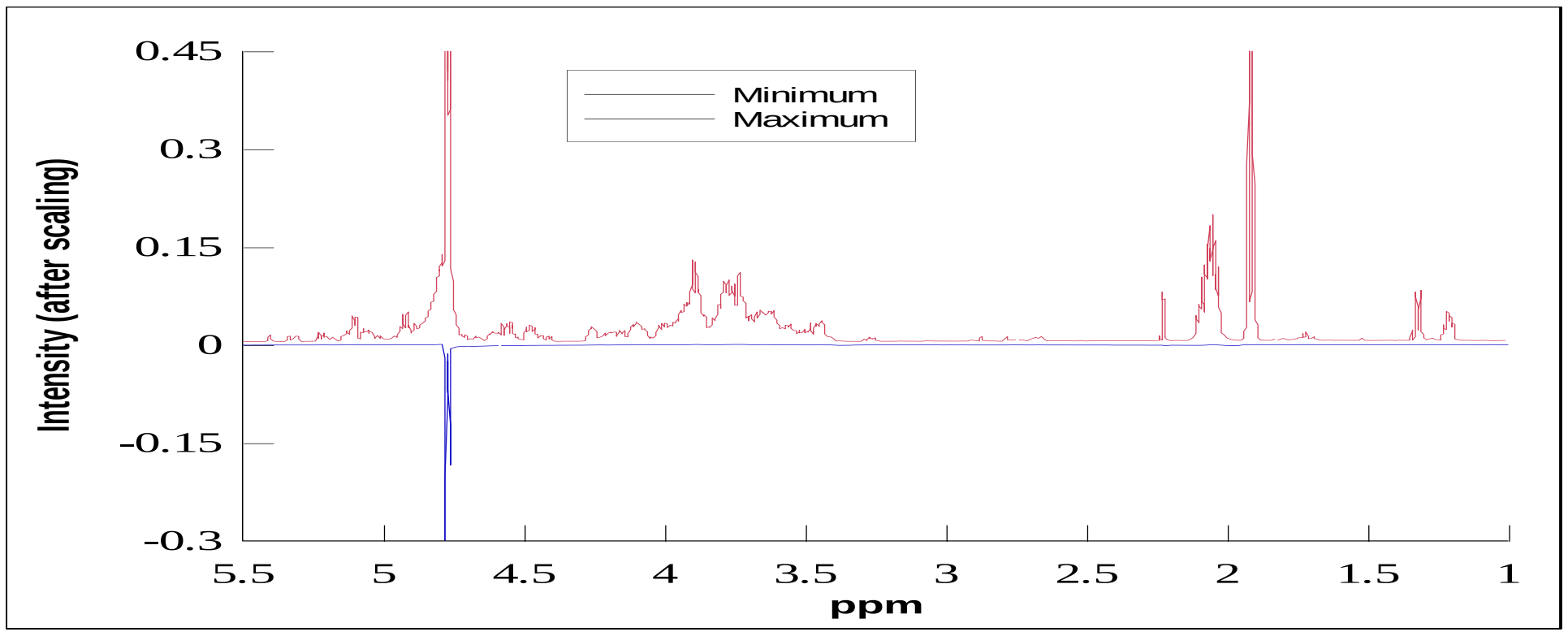

**Figure 1. An example of a minimum and maximum envelope that was constructed using the 80 spectra in the library. The top and bottom curves are the maximum and minimum values discovered in 5000 information channels spanning 5.5-1.0 ppm interval, respectively.**

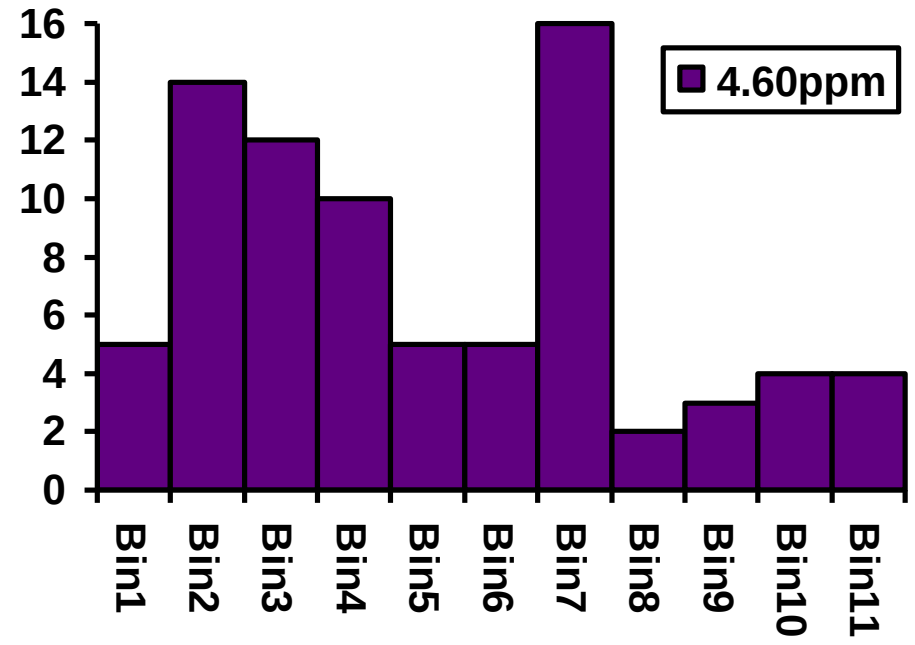

**Figure 2a.**

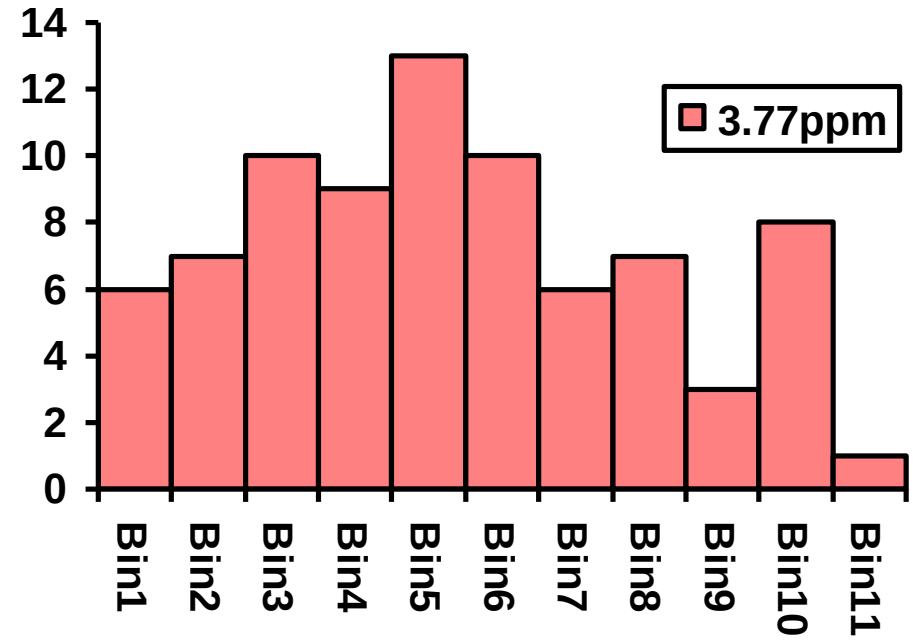

**Figure 2b.**

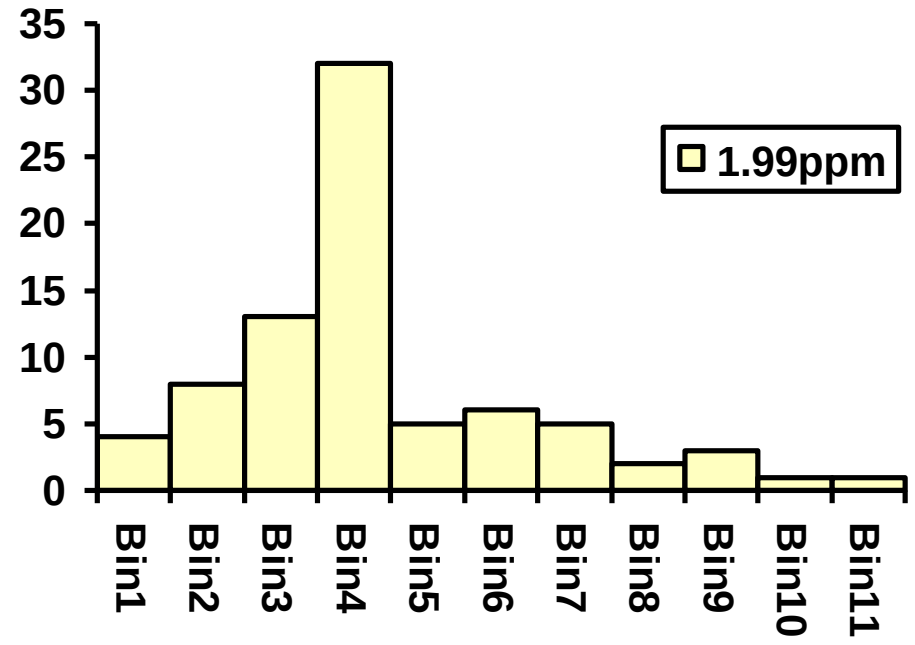

**Figure 2c.**

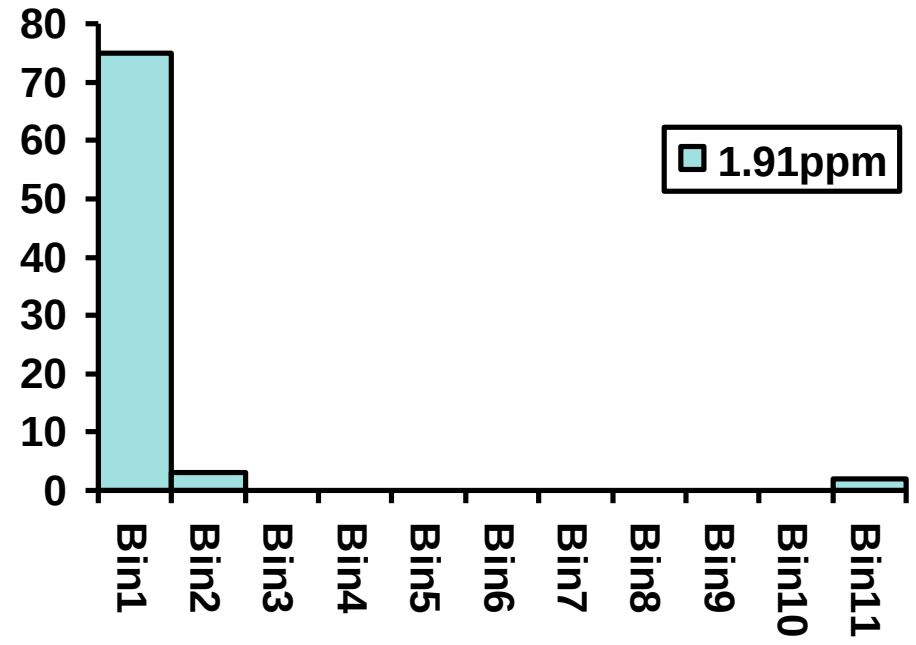

**Figure 2d.**

**Figures 2a-d. Histograms of four different channels illustrating different distributions.**

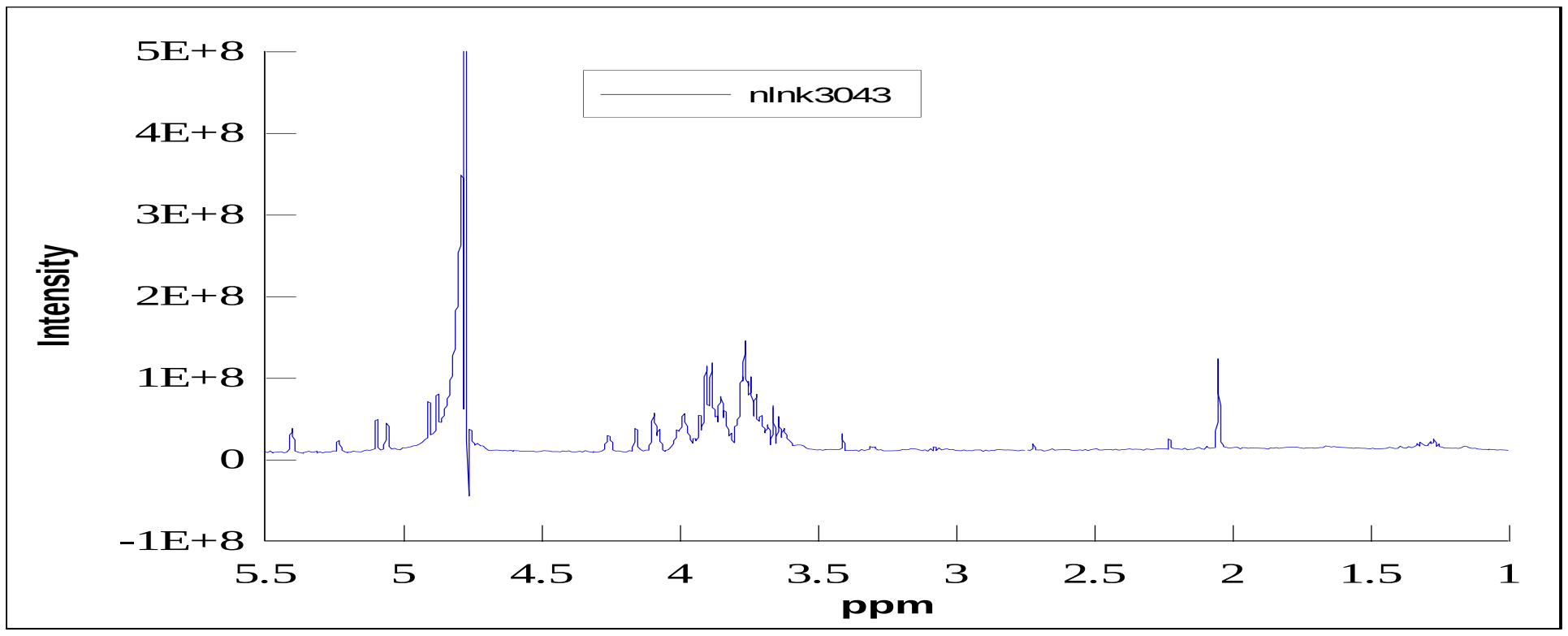

**Figure 3. A sample $^{1}$H-NMR spectrum of an N-linked oligosaccharide identified as 15969 in CarbBank database.**

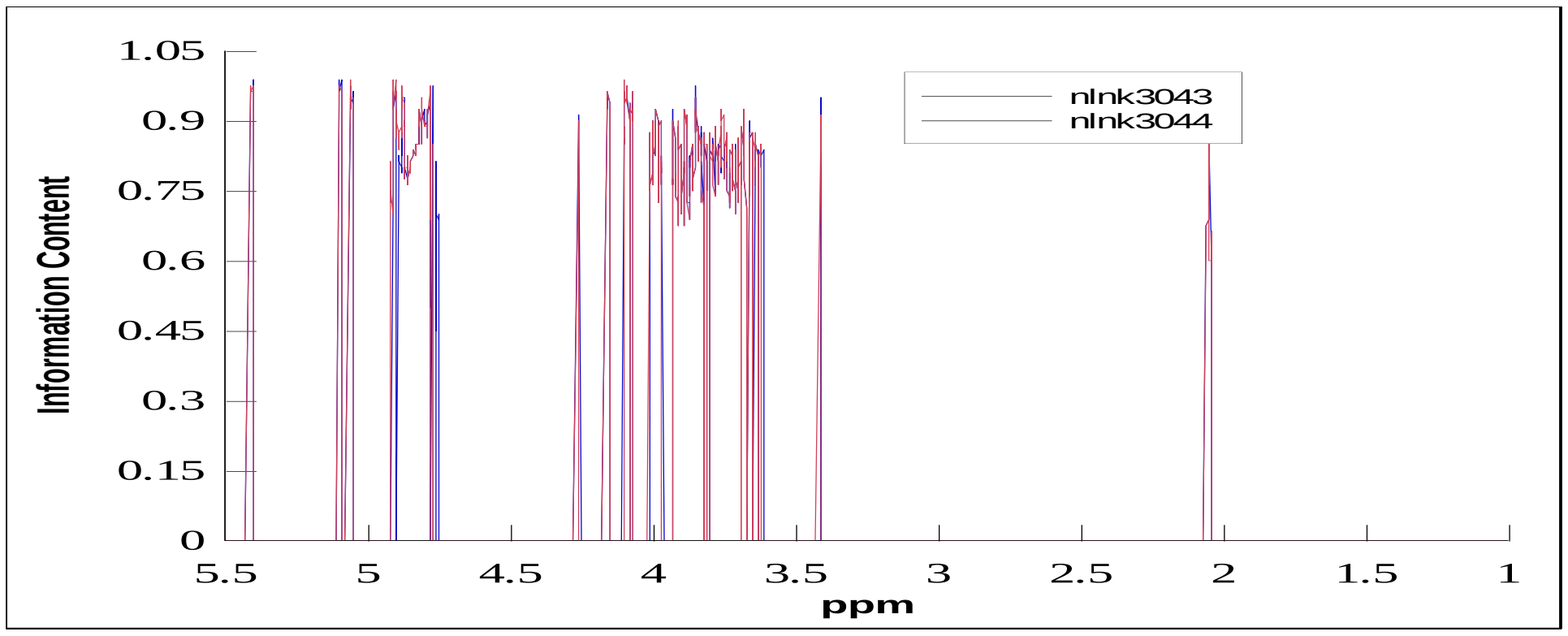

**Figure 4. Two FIS of different spectra of an N-linked oligosaccharide 15969.**

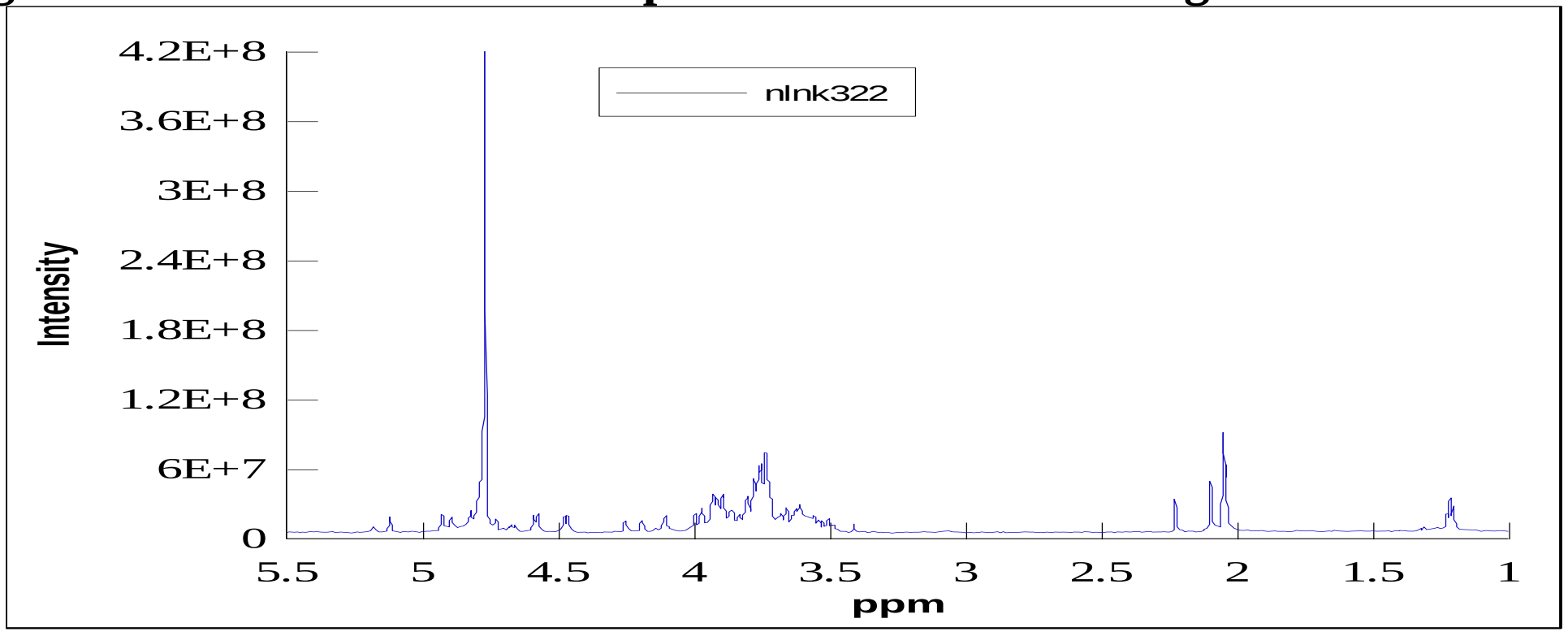

**Figure 5. A sample $^{1}$H-NMR spectrum of an N-linked oligosaccharide identified as 19396 in CarbBank database.**

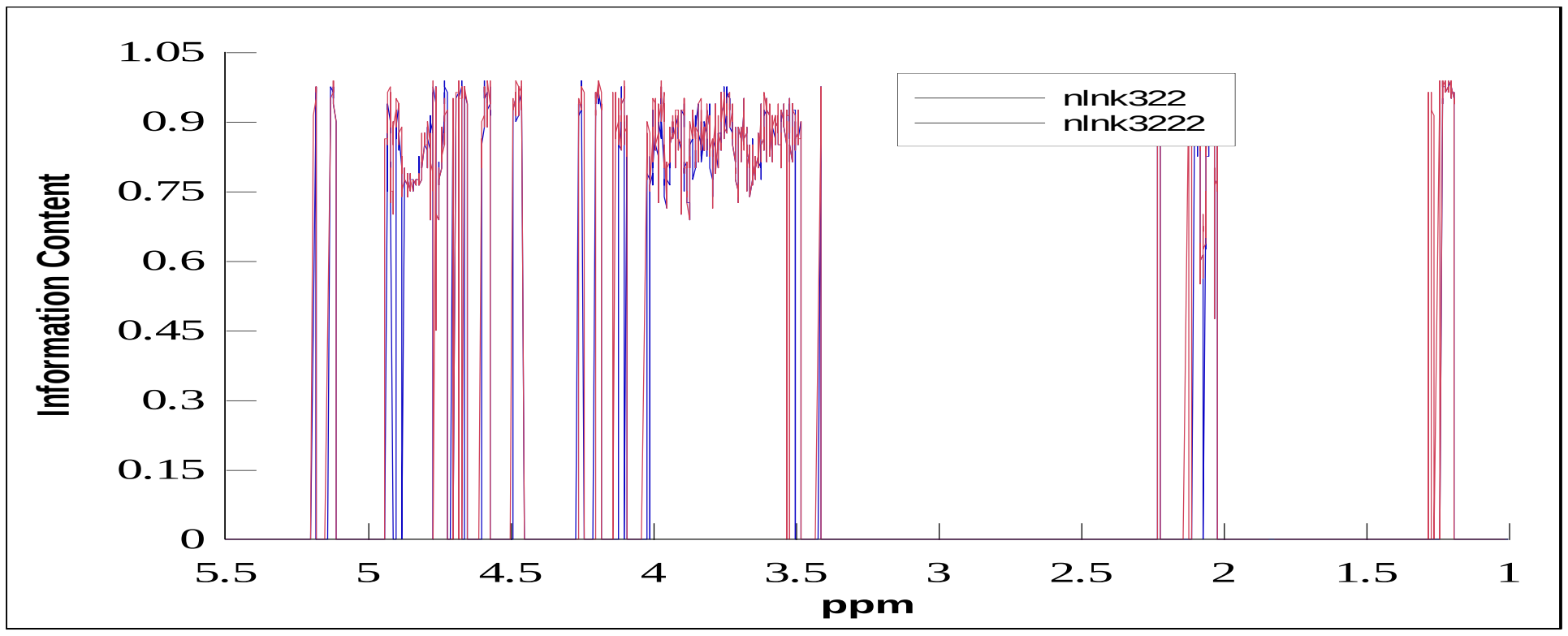

**Figure 6. Two FIS of different spectra of an N-linked oligosaccharide 19396.**

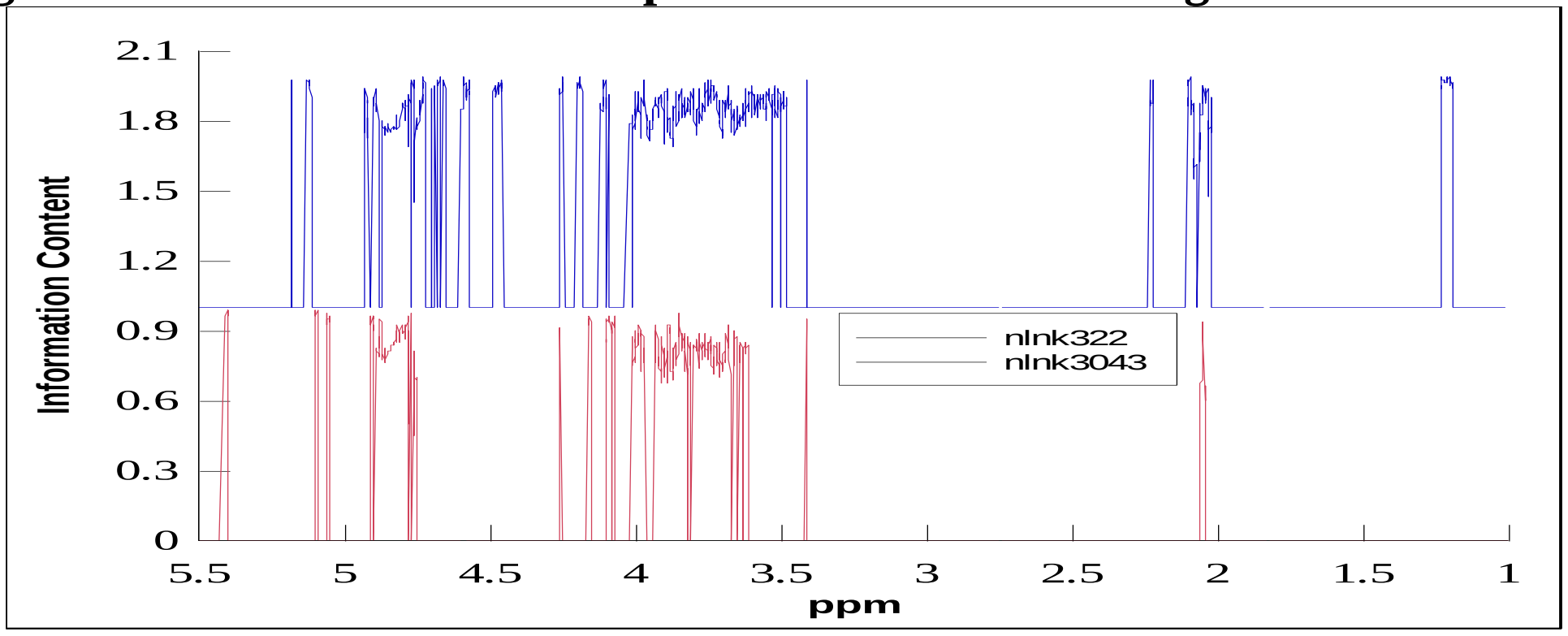

**Figure 7. FIS comparison of two distinct N-linked oligosaccharides 15969 (bottom) and 19396 (top)**

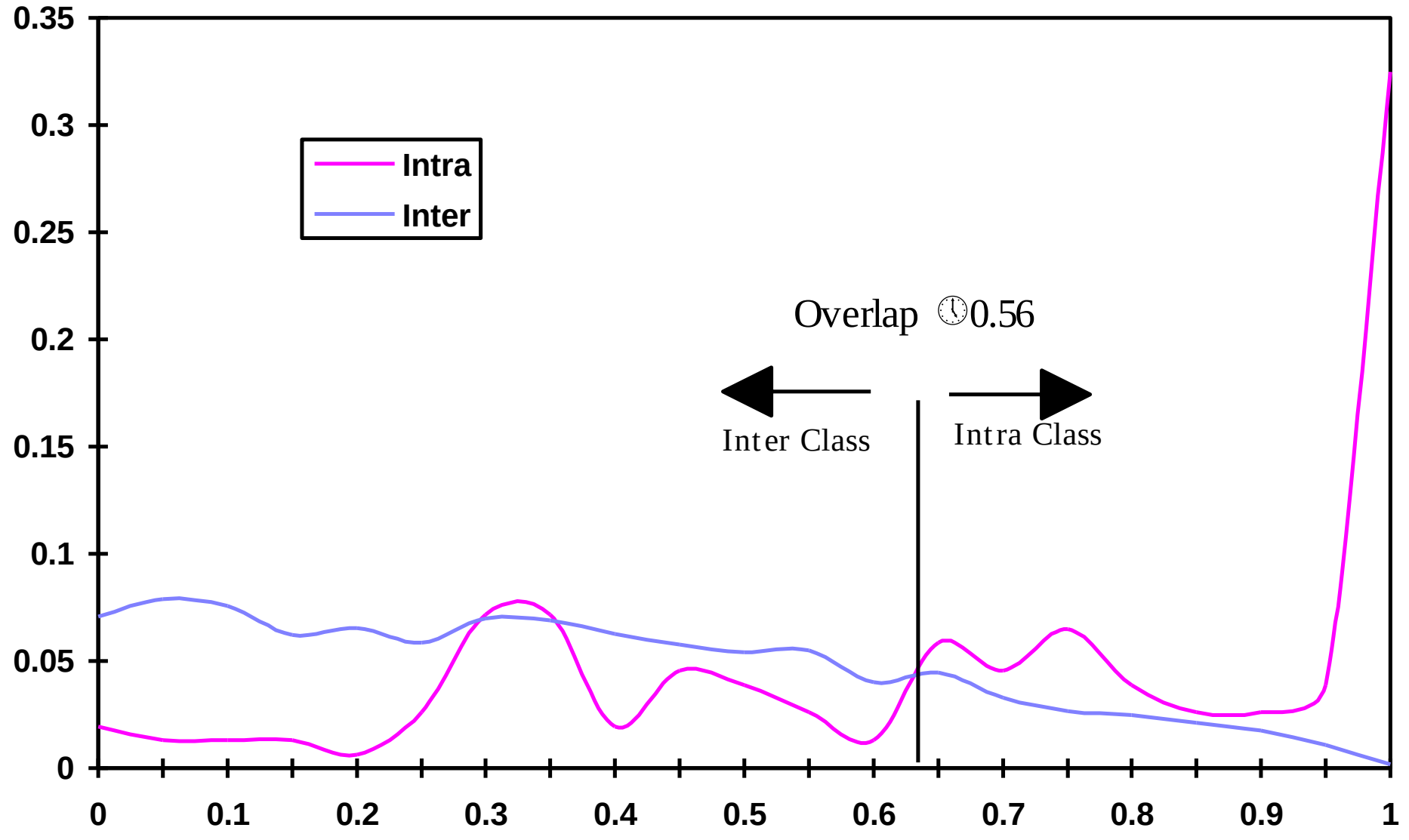

Figure 8(a). Histogram of the correlation coefficient distribution for the raw $^{1}$H-NMR spectra.

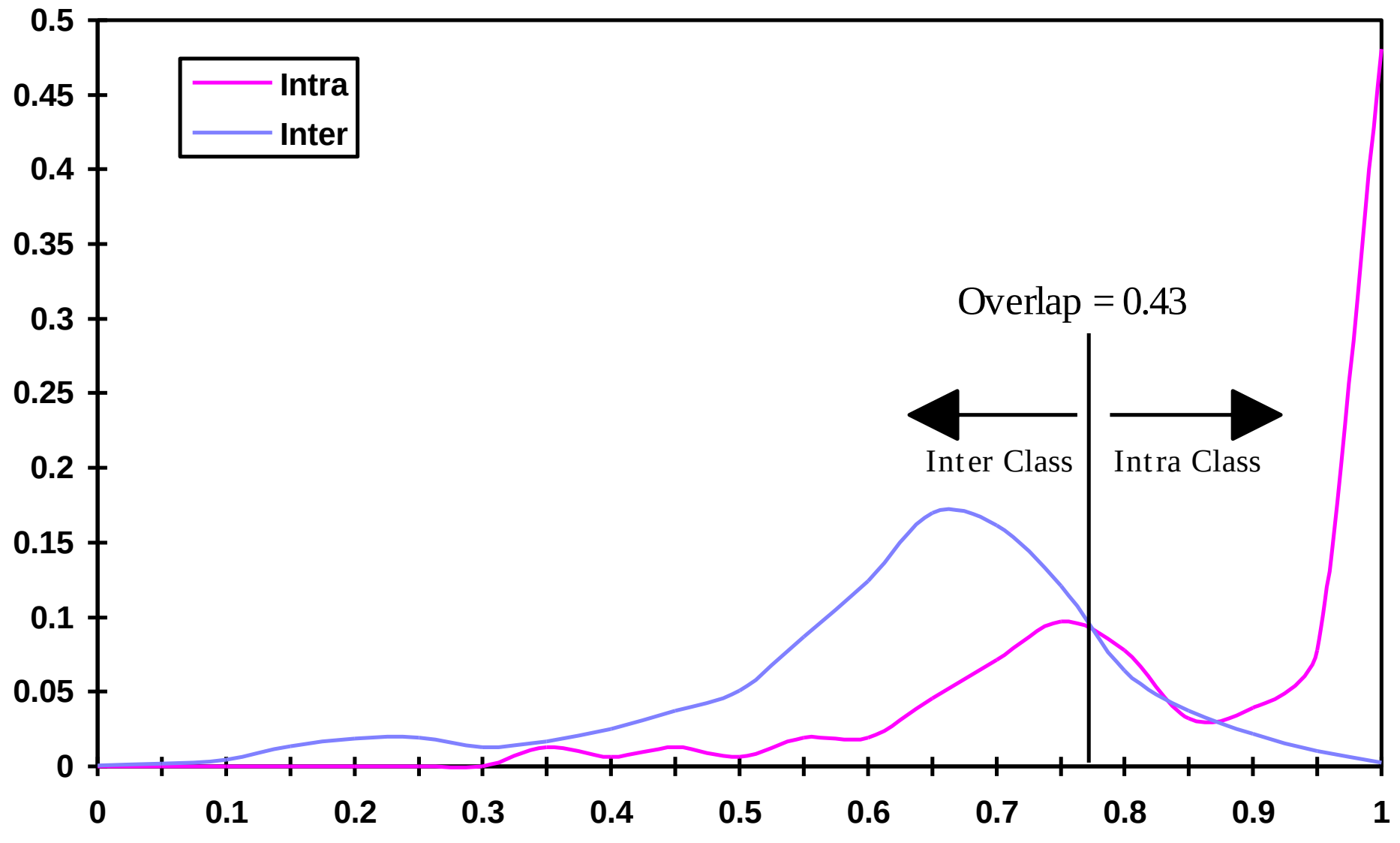

Figure 8(b). Histogram of the correlation coefficient distribution for the SPUTNIK transformed $^{1}$H-NMR spectra.

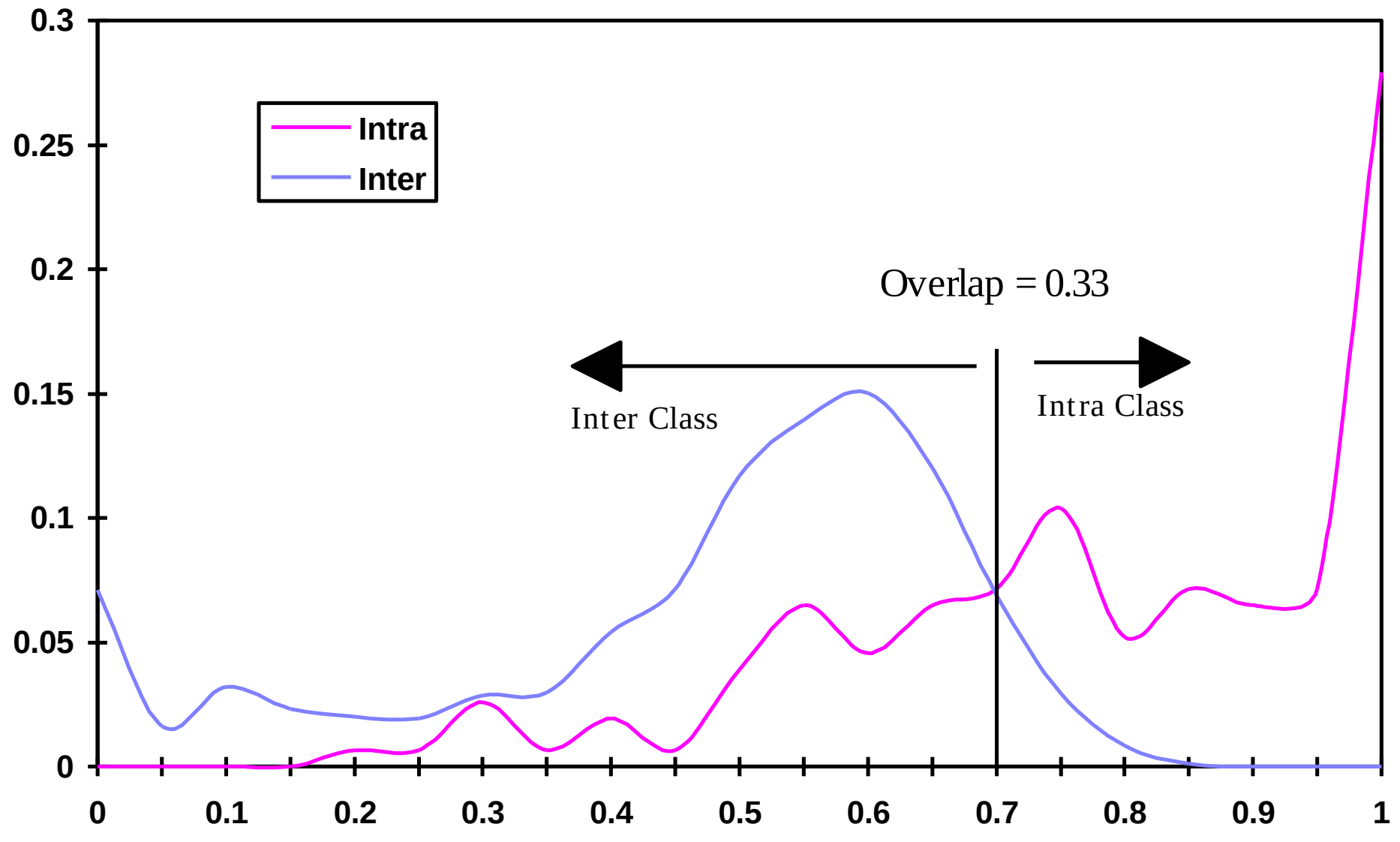

Figure 8(c). Histogram of the correlation coefficient distribution for the FIT transformed $^{1}$H-NMR spectra.

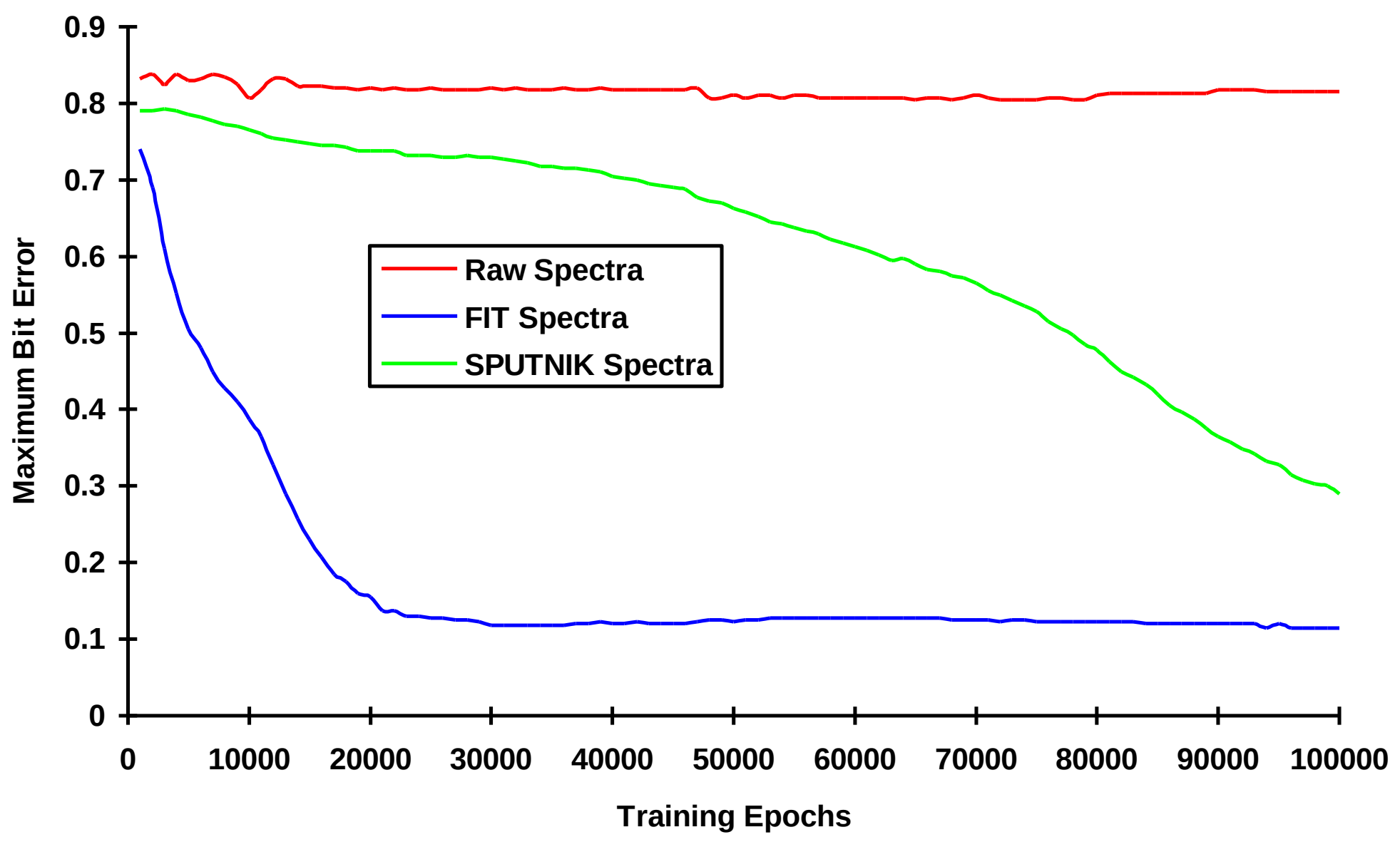

Figure 9. Maximum bit error plotted versus the number of training epochs for raw spectra and FIT spectra.